\documentclass[a4paper,10pt]{article}
\usepackage[utf8]{inputenc}
\usepackage{graphicx} 
\usepackage{amsmath,amssymb}

\title{\textbf{The buckling and invagination process during consolidation of colloidal droplets}\footnote{Soft Matter (2012) DOI: 10.1039/c2sm26530c}}
\author{\textbf{F. Boulogne,\textit{$^{\dagger}$} F. Giorgiutti-Dauphin\'e,\textit{$^{\dagger}$} and
L. Pauchard}\footnote{CNRS, UMR 7608, Lab FAST, Bat 502, Campus Univ - F-91405, Orsay, France, EU. Fax: +33 1 69 15 80 60; Tel: +33 1 69 15 80 49; E-mail: pauchard@fast.u-psud.fr} }

\begin{document}

\maketitle

\begin{abstract}
Drying a droplet of colloidal dispersion can result in complex pattern formation due to both development and deformation of a skin at the drop surface.
The present study focus on the drying process of droplets of colloidal dispersions in a confined geometry where direct observations of the skin thickness are allowed.
During the drying, a buckling process is followed by a single depression growth inside the drop.
The deformation of the droplet is found to be generic and is studied for various colloidal dispersions. 
The final shape can be partly explained by simple energy analysis based on the competition between bending and stretching deformations.
Particularly, the final shape enables us to determine precisely the critical thickness of the shell for buckling.
This study allow us to validate theory in 2D droplets and apply it to the case of 3D droplets where the thickness is not accessible by direct observation.
\end{abstract}

\section{Introduction}

Drying droplets of complex fluids as polymer solutions or colloidal dispersions involves a large number of microscopic phenomena: solvent diffusion, transfers at the vapor/medium interface, skin formation then skin deformation.
Over the last decade, there has been a great deal of scientific and technological interests in studying the flow and deposition of materials in drying droplets \cite{Larson05}.
As an example, the fabrication of dried milk requires the transformation of liquid droplets into a powder form using nozzles; also liquid food concentrate is atomized into droplets using spray-drying processes \cite{Kentish05,Chen09,Handscomb10}.
Particularly functional properties have to be controlled such as dispersibility, and solubility that are essential for the capability of the powder to be re-hydrated.
Therefore, the droplet morphology, or the droplet size changes have to be controlled and the effects of various initial or environmental conditions on a single droplet development have to be understood.

The concentration at the free surface induced by desiccation can cause skin formation of the film \cite{Okuzono06} and is often responsible for strong distortions of the droplet \cite{pauchard03,pauchard11}.
Deformation of nearly spheroidal droplets were considered in the case of polymer solutions\cite{pauchard03} or colloidal dispersions \cite{pauchard04,tsapis05}: the various shapes have been related to buckling process of a skin formed at the drop surface.
Such deformations are not specifically related to drying droplets and can be observed at different scales: capsules formation \cite{Quilliet06,Quilliet12} or lock particles formation when controlled shell buckling that forms spherical cavities\cite{Pine10}.
However in spheroidal structure the thickness of the skin is not directly accessible generally.
Thus, we focus here on a single drop of a colloidal dispersion confined between two circular glass plates and left to evaporate (see figure \ref{fig:setup}).
In this configuration, diffusion processes and drying kinetics have been carefully studied \cite{clement04,leng10,pauchard11,Salmon11,Daubersies12,pauchard12}. 
Moreover this geometry allows us a direct observation and quantification of the skin growth near the drop/air interface by imaging analysis.
When its structural strength is sufficient, the skin can withstand internal stress and can be regarded as an elastic shell.
By appropriate selection of physical quantities, constrained shrinkage can result in buckling instability, causing the circular shell to inversion of curvature as shown in sequence in figure \ref{fig:morpho}a: the resulting single depression grows and is continued by an invagination tube penetrating inside the drop.
The complex pattern formation is studied for various colloidal dispersions and drop sizes.
The generic shape is explained by simple energy analysis.
Particularly, the critical thickness of the shell for buckling is precisely determined by the final shape of the drop.
In addition, this study allow us to validate theory in 2D droplets and apply it to the case of 3D droplets where the thickness is not directly accessible.
The case of three-dimensional droplets and large P\'eclet numbers corresponds to common industrial processes (spray-drying consisting in rapidly drying aerosols to manufacture dry powders).

\section{Experimental}

\begin{figure}
\centering
\resizebox{0.55\columnwidth}{!}{
  \includegraphics{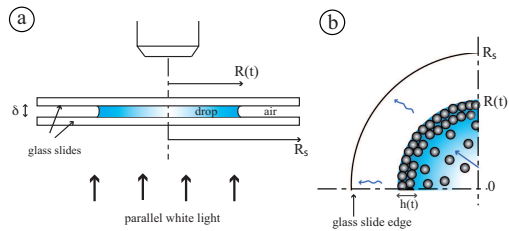}
}
\caption{(a) Set-up: a droplet of solution is sandwiched between two circular glass slides in side view (left); the cell is illuminated by transmitted light. (b) Sketch in top view of a quarter of cell showing the formation of a densely-packed particles at the liquid-vapour interface during solvent removal.}
\label{fig:setup}       
\end{figure}

Our experiments were performed with various stable aqueous dispersions (see Table \ref{t.1}).
(i) silica particles Ludox HS-40 (solid volume fraction $\phi=0.22$, particle diameter $2a=15\pm2$ nm, polydispersity index $\sim 0.16$) and diluted dispersion of Ludox TM-50 (solid volume fraction $\phi=0.12$, particle diameter $2a=22\pm2$ nm, polydispersity index $\sim 0.2$ \cite{Triolo87}) from Sigma-Aldrich;
(ii) Nanolatex particles, polystyrene (solid volume fraction $\phi=0.25$, particle diameter $2a=25$ nm, glass transition temperature $=100^\circ$C) provided by Rhodia Recherche (Aubervilliers, France);
(iii) Green fluorescent silica particles sicastar-greenF (solid volume fraction $\phi=0.10$, particle diameter $2a=50\pm3$ nm, polydispersity index $\sim 0.2$, absorbance: $\lambda_{excitation} = 475$ nm and emission: $\lambda_{emission} = 510$ nm) commercially available from Micromod.
Most of the experiments were performed with the last system.
The colloidal particles are adequately polydispersed to avoid cristallization. 
In addition during the drying process, the particles density increases resulting in the formation of a gel phase, defined as a porous matrix saturated with water.
Dilution of dispersion was possibly done by adding deionized water (quality Milli-$\rho$).
The confined geometry consists of a thin cell made of two circular, parallel and horizontal glass slides of radius $R_{s}$ (figure \ref{fig:h}).
It has been shown that the drying kinetics can be controlled by the slide size and is not strongly affected by the relative humidity of the surrounding\cite{clement04};
experiments were conducted at room temperature and relative humidity ($\sim 40\%$).
The slides are lubricated with a thin film of silicon oil $V1000$ (viscosity $1$ Pa.s).
This treatment prevents from the pinning process of the contact line \cite{pauchard11}.
A drop of the dispersion is placed on the bottom substrate with a micropipet.
Then the upper glass slide is carefully placed to squeeze the drop (see figure \ref{fig:setup}a). 
The gap $\delta$, between the slides is controlled using three thin spacers of controlled thickness.
In all experiments the radius of the glass slides is kept to $R_{s} = 8$ mm and the gap is constant, $\delta = 100 \pm 5 \mu$m.
During the drying process, images of the drop evolution are recorded at different times and show a dark ring corresponding to the meniscus at its periphery as shown in figure \ref{fig:morpho}a.
The high contrast of the images allows an accurate detection of the drop edge and provides variations with time of both the periphery length and surface area of the drop.
Fluorescent microscopy was performed using a DM2500 Leica microscope with objective $5\times$ and $50\times$ magnification for global view (drops deformation) and close-up view (envelope thickness measurements) respectively. 
The measurement of the size of the densely-packed region at the drop periphery, envelope thickness, have been investigated by fluorescence microscopy (see figure \ref{fig:shell}).
Focusing on the air/drop interface, in the middle of the cell ($z=\frac{\delta}{2}$ in figure \ref{fig:shell}a), the size of the envelope thickness, $h$, can be estimated from the fluorescence intensity profile with an accuracy of $5\mu$m (figure \ref{fig:shell}b). 

\section{Results}
\label{sec:3}

The deformation of the drop during the drying process exhibits different configurations shown in the sequence in figure \ref{fig:morpho}a and in the superposition of the drop profiles in figure \ref{fig:morpho}b.
The successive configurations adopted by the drop are driven by the volume decrease that is induced by solvent removal.
In a first stage the drop progressively shrinks with solvent evaporation (figure \ref{fig:morpho}a(1),(2)). 
Then, contrasting with the case of a pure solvent, the droplet stops shrinking isotropically: a sudden inversion of curvature occurs (figure \ref{fig:morpho}a(3)).
Consequently, the depression is continued by an invagination tip that deepens with time (figure \ref{fig:morpho}a(4),(5)) until the formation of a bean-shaped material that cracks in figure \ref{fig:morpho}a(6)).
These configurations are detailed in the following.
Most of quantities used are shown in figure \ref{fig:h}.

\subsection{Isotropic shrinkage and shell formation}

\begin{figure}
\centering
\resizebox{0.55\columnwidth}{!}{
  \includegraphics{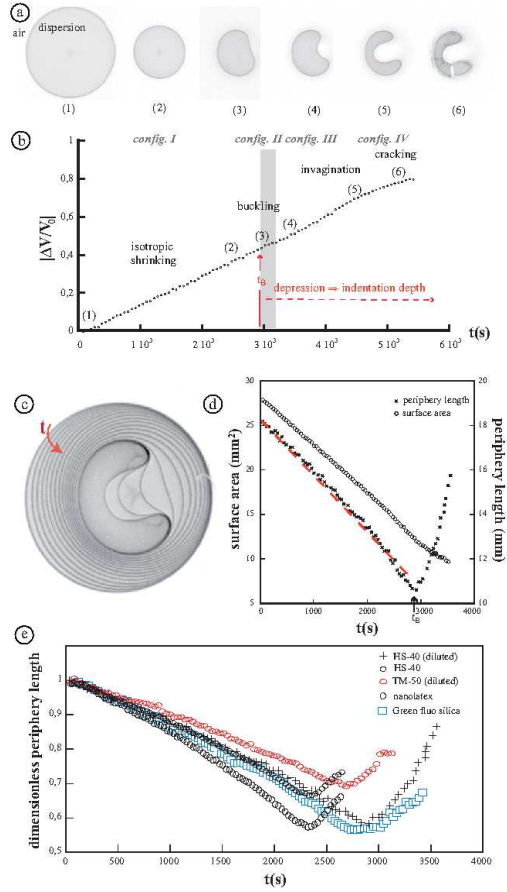}
}
\caption{(a) Digitized images taken during a typical deformation of a drop of colloidal dispersion (Ludox HS-40 (diluted), $R_{0}=2.9$mm, $R_{s}=8$mm). (b) Time variation of the relative volume of the drop (measurements); at the buckling onset a indentation depth, can be defined and characterize the growth of a depression. (c) Superposition of drop profiles during the drying process: duration elapsed between two images is $1$ minute. (d) Time variations of periphery length (crosses) and surface area (circles) of the drop; the inversion of curvature takes place at time $t_{B}$, says the buckling time. The dashed line corresponds to the linear decrease of the drop periphery length with time. (e) Time variations of the dimensionless periphery length for various drops.}
\label{fig:morpho}       
\end{figure}

During the drying process, the gas that escapes from the liquid/air interface is driven outward from the cell by diffusion \cite{clement04}.
Direct observation of the drop evolution with time shows a progressive and isotropic shrinkage in a first stage (figure \ref{fig:morpho}a (from (1) to (2)), figure \ref{fig:morpho}b): this first stage is named as configuration I.
As a result both periphery length and surface area decrease steadily with time\cite{pauchard11} (figure \ref{fig:morpho}d).
In the following we suppose that the radius decreases linearly with time as: $R(t) = R_{0}-Jt$ where $R_{0}$ is the initial drop radius and $J$ is the evaporation rate.
In addition time variations of dimensionless periphery length are shown in figure \ref{fig:morpho}e for different drops.
In particular, the duration of the shrinkage process increases when the initial volume fraction of dispersions dereases (this is observed in the case of pure and diluted Ludox HS-40).

\begin{figure}
\centering
\resizebox{0.55\columnwidth}{!}{
  \includegraphics{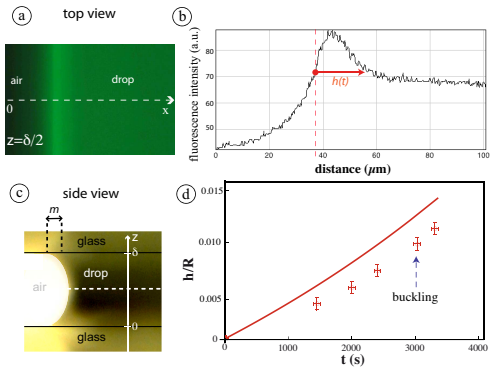}
}
\caption{(a) Fluorescence images in top view (dispersion of Green fluorescent silica particles). (b) Corresponding intensity profiles across the air/drop interface, along the dashed line in (a); in the referential bound to the drop/air interface, the increase of the envelope thickness $h(t)$, can be measured with time. (c) Menisci at the drop edge in side view $z=0$ and $z=\delta$ denote positions with the substrates ($\delta=100 \pm 5\mu$m); the contact angle of the aqueous dispersion on the surface is close to $60^o$. (d) Ratio of the envelope thickness $h$, to the drop radius $R$, as a function of time: crosses are measurements deduced from the intensity profiles in (b) ($R_{0} = 2.9$mm, $R_{s} = 8$mm), dashed line is theory in accordance with expression \ref{e.1} in the isotropic regime.}
\label{fig:shell}       
\end{figure}

Starting from suspended compounds uniformly distributed within the droplet, particles are advected to the drop periphery due to solvent removal (figure \ref{fig:setup}b).
The radial flow, induced by the evaporation rate, gives the typical velocity scale characterizing the transport of particles.
A P\'eclet number can be defined as the diffusion time $t_{D} = \frac{R_{0}^2}{D_{0}}$ (characteristic time for the diffusion of a particle along a distance $R_{0}$) divided by the evaporation time $t_{E} = \frac{R_{s}}{J}$, as follows:
$Pe = \frac{R_{0}^2 J}{R_{s} D_{0}}$
with the diffusion coefficient $D_{0} = \frac{k_{B}T}{6\pi \eta a} = 8.5 \times 10^{-12} m^2.s^{-1}$ using the Stoke Einstein relation ($\eta$ is the viscosity of the suspending fluid) and the evaporation deduced from the time variation of the periphery length, $J \sim 10^{-7}m.s^{-1}$.
In our experimental conditions: $10 < Pe < 10^2$, diffusion is slow compared with the rate of evaporation. 
As a result, the transport of particles to the drop-air interface strongly suggests the formation of a porous envelope during the evaporation process \cite{Keddie10,Style2011}.
This envelope thickens with time as shown by the direct measurements in figure \ref{fig:shell}d.

Due to the drop shrinkage, the structure of the envelope is possibly modified by a redistribution of particles. 
In this hypothesis, the particle volume fraction in the envelope increases until a maximum value close to the random close packing $\phi_{c}$.
Applying mass conservation of particles the relative thickness $h$, of the shell of radius $R$ can be determined during isotropic shrinkage\cite{tsapis05}.

\begin{equation}
\label{e.1}
\frac{h}{R} = 1-\left(\frac{\phi_c-\phi \left(\frac{R_{0}}{R}\right)^2}{\phi_c-\phi}\right)^{1/2}
\end{equation}

Figure \ref{e.1} allows us to plot the time variation of the envelope thickness from the time variation of the drop radius.
Measurements of the envelope thickness are well fitted by this expression in the isotropic regime (figure \ref{fig:shell}d).
Due to the meniscus the thickness is possibly underestimated which could explain the discrepancy with expresion \ref{e.1}.
The thickening of the envelope does not impede solvent loss since the transport takes place through the envelope in accordance with the Darcy law:

\begin{equation}
\label{e.2}
J = \frac{k}{\eta}\frac{\Delta P}{h}
\end{equation}

where $\Delta P$ is the pressure drop across the shell thickness $h$, required to produce a given fluid flow and $k$ is the permeability of the porous shell estimated using the Carman-Kozeny relation\cite{Brinker90}.

\begin{figure}
\centering
\resizebox{.9\columnwidth}{!}{
  \includegraphics{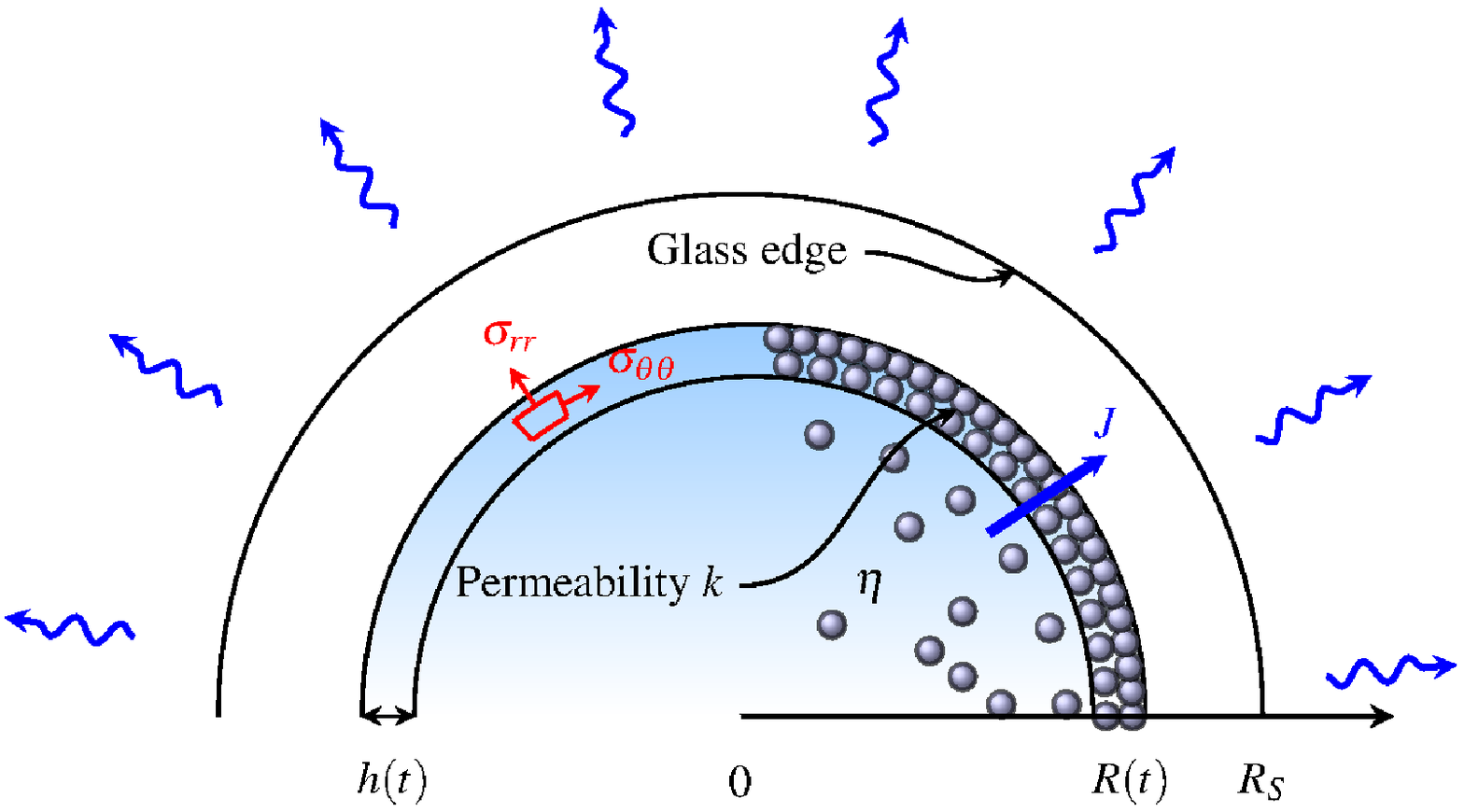}
}
\caption{Mechanical stress field in cylindrical coordinates (left part of the sketch) and definition of physical quantities in the drying droplet (right part of the sketch).}
\label{fig:h}       
\end{figure}

When the structural strength of the envelope is sufficient, the envelope can withstand internal stress.
At this time, particles do not redistribute in the envelope that will be considered as an elastic shell.
The material is assumed to be homogeneous and characterized by an elastic modulus $E$; $E$ is supposed to be constant during the deformation since the gel is continuously drained by the solvent.

\begin{figure}
\centering
\resizebox{0.55\columnwidth}{!}{
  \includegraphics{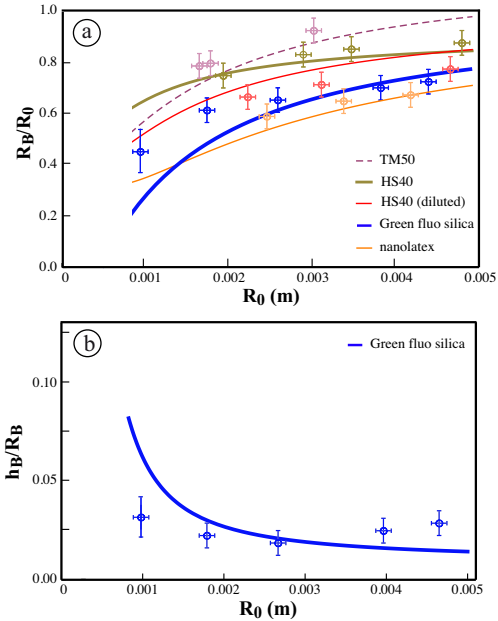}
}
\caption{(a) Ratio of the radius of the shell at the onset of buckling to the initial drop radius $\frac{R_{B}}{R_{0}}$, as a function of $R_{0}$.
(b) Ratio of the critical envelope thickness to the drop radius at the point of buckling $\frac{h_{B}}{R_{B}}$, as a function of $R_{0}$. Dots are measurements and lines are fits to the theoretical expressions \ref{e.1} and \ref{e.3}.}
\label{fig:buckling}       
\end{figure}

In addition the mechanical stress is assumed to be homogeneously distributed in the elastic shell.
Observing that the shell thickness is much smaller than the other two spatial lengths (drop radius $R$ and cell gap $\delta$), the following assumptions can be made.
Let us consider the three components of the mechanical stress field in the elastic shell, in cylindrical coordinates: the ortho-radial, radial and normal components of the stress field are denoted by $\sigma_{\theta\theta}$, $\sigma_{rr}$ and $\sigma_{zz}$ respectively (see figure \ref{fig:h}).
Equilibrium of the corresponding internal forces in the elastic shell results in the following relations:
$\frac{\sigma_{\theta\theta}}{\sigma_{rr}} \sim \frac{R}{h} >> 1$ and $\frac{\sigma_{\theta\theta}}{\sigma_{zz}} \sim \frac{R}{\delta} >> 1$  since $\frac{h}{R} << 1$ during the elastic deformation of the envelope.
Therefore the main component of the stress in the shell is the ortho-radial one. 
This quantity is responsible for the buckling process of the shell.
Indeed, above a critical thickness, deformations occur mainly by bending, which is much less energetic than stretching.
Therefore a buckling process occurs leading to an inversion of curvature of the shell (figure \ref{fig:buckling}a and figure \ref{fig:buckling}b).
In the following suffix $B$ is related to quantities taken at time $t_{B}$.
Expression \ref{e.2} gives the relation between the shell thickness $h_{B}$, and the pressure drop $\Delta P_{B}$ across $h_{B}$ at the onset of buckling:

\begin{equation}
\label{e.3}
h_{B}=\frac{\Delta P_{B}}{J}\frac{k}{\eta}
\end{equation}
 
Firstly expression \ref{e.1} at $t=t_{B}$ gives $R_{B}$ as a function of $R_{0}$ and $h_{B}$.
Replacing $h_{B}$ by the expression \ref{e.3}, $\frac{R_{B}}{R_{0}}$ can be expressed as a function of $R_{0}$ and the pressure drop across the shell at the onset of buckling, $\Delta P_{B}$. In this way $\Delta P_{B}$ is a parameter and can be estimated for each system studied.
$\frac{R_{B}}{R_{0}}$ slightly increases with the initial drop radius as shown by the measurements for different colloidal systems (figure \ref{fig:buckling}a).
These results are close to the linear increase of $R_{B}$ with $R_{0}$ obtained in spherical drops by Tsapis \textit{et al.}\cite{tsapis05}.
Secondly, combining expressions \ref{e.1} and \ref{e.3} the critical thickness at the point of buckling is plotted in figure \ref{fig:buckling}b; measurements of the thickness using the method described in figure \ref{fig:shell}.
The order of magnitude of $\Delta P_{B}$ is found to be in agreement with the values obtained by Tsapis \textit{et al.}\cite{tsapis05}. 
Values are listed in Table \ref{t.1} for different systems.

\begin{table}
\caption{Main characteristics for the samples considered in the experiments: particle diameter $2a$ (and polydispersity is known), particle volume fraction $\phi$, permeability of the gel $k$, and critical pressure drop across the shell at the onset of buckling $\Delta P_{B}$, deduced from the measurements.}
\label{t.1}
\begin{center}
\begin{tabular}{ccccc}

& $2a (nm)$ & $\phi$ & $k (\times 10^{-18} m^2) $ & $\Delta P_{B} (kPa)$\\
\hline
Ludox HS-40 & $15  \pm 2$ & $0.22$ & $1.2$ & $9$\\
Ludox HS-40 (diluted) & $15  \pm 2$ & $0.10$ & $1.2$ & $9$\\
Ludox TM-50 (diluted) & $22 \pm 2$ & $0.12$ & $2.4$ & $5$\\
nanolatex & $25$ & $0.25$ & $4.4$ & $3$\\
Green fluo silica & $50 \pm 3$ & $0.10$ & $12$ & $0.8$\\
\end{tabular}
\end{center}
\end{table}

\subsection{Depression growth}

\begin{figure}
\centering
\resizebox{0.55\columnwidth}{!}{
  \includegraphics{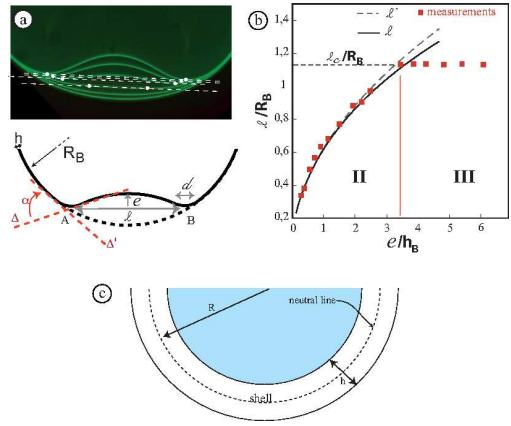}
}
\caption{(a) Top: superposition of digitized profiles showing the depression growth in a part of the shell; the whites dots locate the regions of highest curvature. Down: sketch of the depression formed by inversion of the circular cap: the indentation length $e$, happens in a region of size $\ell$, between $A$ and $B$. The angle $\alpha$ between the asymptotes $\Delta$ and $\Delta'$ limit the fold; the quantity $d$ is the lateral extension of the fold limiting the circular shell and the inverted cap. (b) Ratio of the lateral size of the depression, $\ell_{exp}$, to the radius of the drop at the onset of buckling $R_{B}$, as a function of the indentation length $e$, to the critical thickness $h_{B}$, at the onset of buckling; $\ell' = 2 \sqrt{2R_B e-e^2}$, $\ell = 2 \sqrt{2R_B e}$. $h_{B} = 12\pm5\mu m$. Above a critical indentation length, the quantity $\ell$ becomes constant: $\ell=\ell_{c}$. 
(c) Sketch of the shell forming at the drop periphery: the geometrical quantities of the shell are defined taking into account the shell thickness $h$.}
\label{fig:energies}       
\end{figure}

As shown previously, the shell stops shrinking ($R=R_{B}$) and a buckling process results in the formation of a depression in a region of the shell (figure \ref{fig:morpho}b and figure \ref{fig:energies}a).
In the following, the existence of the depression in the shell is referred as configuration II.
A part of the elastic energy is released by the formation of this depression (figure \ref{fig:energies}a).
The depression growth is characterized by the increase of the indentation length $e$, and happens in a region of linear size $\ell$.
Measurements of $\ell$ as a function of $e$ are shown in figure \ref{fig:energies}b. 
These results are well fitted by the geometric relation: $\ell = 2 \sqrt{2R_B e-e^2}$ approximated by $2 \sqrt{2R_B e}$ to first order in $e$, as a consequence of the inversion of curvature.

The elastic energy of configuration II is the sum of three contributions:

\begin{equation}
\label{e.4}
U_{II} =  U_{0}+U_{fold}+U_{inversion}
\end{equation}

\begin{enumerate}

\item The term $U_{0}$ is the contribution of the elastic energy due to the part of the shell not inverted: it is considered as a reference energy.

\item During the depression growth the elastic energy contains a main contribution coming from the creation of the folds $A$ and $B$ (see figure \ref{fig:energies}a): a fold is limited by asymptotes (asymptotes $\Delta$ and $\Delta'$ for $A$) and extends over a typical distance $d$.
Assuming $h/R<<1$, which is actually realized during the drop deformation, the F\"oppl-von K\`arm\`an theory for elastic shells \cite{Foppl07,Landau86,Benamar97} allows us to obtain the energy due to the fold $U_{fold}$.

Minimizing the total elastic energy (bending energy and in-plane elastic energy) with respect to the distance $d$ leads to $d = (h R_B)^{1/2}$ (this was described in references \cite{Landau86,pauchard97}).
Consequently the relevant curvature radius is $d/\tan{\alpha}$ where $\alpha$ is the angle between the asymptotes limiting the fold (figure \ref{fig:energies}a).
Therefore $U_{fold} \sim E \frac{h^2 s}{R_{B}} \tan{\alpha/2}^2$, where $s=2d \delta$ is the area of the fold\cite{pauchard97,pauchard98}.
As the depression grows, the angle $\alpha$ increases, in accordance with $\tan{\alpha/2}^2 \sim \frac{e}{R_{B}}$, so does the energy $U_{fold}$.
$U_{fold}$ is due to both the increase of the indentation length and the increase of the shell thickness.
Consequently, at each time denoted by $t_{i}$, an indentation length $e_{i}$, and a shell thickness $h_{i}$, can be defined. 

\begin{equation}
\label{e.5}
U_{fold}(h_{i},e_{i}) = 2c_{0} \delta E h_{i}^{5/2} \frac{1}{R_{B}^{3/2}} e_{i}
\end{equation}

where $c_{0}$ is a parameter that only depends on the Poisson ratio of the shell.

\item The term $U_{inversion}$ corresponds to the change in elastic energy due the inversion of the circular section $AB$ in accordance with figure \ref{fig:energies}c.
In this way the arc of external radius $R_{B}+h_{B}/2$ becomes compressed after inversion; also the arc of internal radius $R_{B}-h_{B}/2$ becomes stretched after inversion.
The length change makes this energy contribution linear in the deformation \cite{pauchard98}.

\begin{equation}
\label{e.6}
U_{inversion}(h_{i},e_{i}) = \frac{c_{1}\sqrt{2}}{4} E \delta h_{i}^{3} \frac{1}{R_{B}^{3/2}} e_{i}^{1/2}
\end{equation}

where $c_{1}$ is a parameter that only depends on the Poisson ratio of the shell.
Note that $U_{inversion}(h_{i},e_{i})$  becomes negligible compared with $U_{fold}(h_{i},e_{i})$ for $e_{i}>>h$.

\end{enumerate}

\begin{figure}
\centering
\resizebox{0.55\columnwidth}{!}{
  \includegraphics{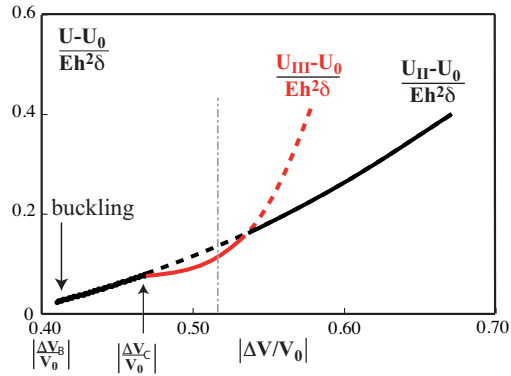}
}
\caption{Dimensionless elastic energy as a function of the relative volume variation $\mid \frac{\Delta V}{V_{0}} \mid$ (the lowest-energy state is plotted as a solid line and the higher-energy one is plotted as a dashed line); the plotrange starts at the buckling process $\mid \frac{\Delta V_{B}}{V_{0}} \mid=\frac{R_{B}^2}{R_{0}^2}$. At a threshold value, $\mid \frac{\Delta V_{C}}{V_{0}} \mid$, configuration II becomes energetically more favourable than configuration III. The high stretching process occuring during the configuration III possibly causes weakening and breakage of the inverted part of the shell; in a realistic point of view the vertical dashed line is not reached during the deformation of the elastic shell.}
\label{fig:energy}       
\end{figure}

The energy $(U_{II}-U_{0})/(Eh^2\delta)$ is plotted as a function of the relative volume variation, $\left| \frac{\Delta V}{V_{0}} \right| \sim \frac{\pi R_{B}^2-(2R_{B})^{1/2}e^{3/2}}{\pi R_{0}^2}$ in figure \ref{fig:energy}. 
This illustrates the tendency of the elastic energy to increase during the shell deformation.

Since the energy concentrated in the folds becomes enormous during the depression growth, another configuration, denoted in the following by configuration III, is energetically more favourable.
This is realized by assuming the angle $\alpha$ constant ($\alpha$ defined in figure \ref{fig:energies}a).
Hence, the lateral size of the depression $\ell$, stops increasing and becomes constant as shown by our measurements in figure \ref{fig:energies}b (region named III). 
Now, the elastic energy due to the fold only increases through the growth of the shell thickness.

However, the decrease of the inner volume of the shell need another mode of deformation of the shell.
Assuming $\alpha$ constant and keeping $\ell$ constant, as shown by empirical observations, require an increase of the indentation length $e$ (figure \ref{fig:energies}a) and results in a length change of the inverted part of the shell between $A$ and $B$.
Let us consider the stretching energy due to the length change of the inverted part of the shell.
Usually, when an elastic material is deformed the work is stored in the body as a strain energy.
If the elastic deformation $\epsilon'$ (strain), occurs in an elementary volume $dV \sim \delta h \ell$ ($\ell$ being the length scale of the stretched region), the elastic energy expresses as:

$$U_{stretch} = E \int dV \epsilon'^2$$

As a result, the energy of the inverted part due to the relative change in length $u$ expresses as:

\begin{equation}
\label{e.7}
U_{stretch}(h_{i},e_{i}) =  c_{2}\delta E h_{i} \ell u(e_{i})^2
\end{equation}

where $c_{2}$ is a parameter that only depends on the Poisson ratio of the shell.
The relative length change $u(e_{i})$ differs from zero only for $e_{i} > e_{c}$.
Then, the length change is related to the indentation length by\footnote{Starting from a circular arc AB, the simplest perturbation corresponding to an elongational shape is a part of an ellipse.
We use the approximation by Ramanujan \cite{Barnard01} for the circumference of an ellipse gives:
$C_{(a,b)} \sim \pi \big(3(a+b)-\sqrt{(3a+b)(a+3b)} \big) $ where $a$ and $b$ are one-half of the ellipse's major and minor axes respectively.
Taking $a=R_{B}+e_{i}$ and $b=R_{B}$, the relative length change in length between such an ellipse and a circle of radius $R_{B}$ expresses as: $\frac{C_{(R_{B}+e_{i},R_{B})}-C_{(R_{B},R_{B})}}{C_{(R_{B},R_{B})}}  \sim \frac{e_{i}}{2R_{B}}$ to first order in $e_{i}$. The relative length change related to the transition from configuration II to configuration III in \ref{fig:energy} is then $\frac{e_{i}}{4R_{B}}$.}
$u(e_{i}) \sim \frac{1}{2}\frac{e_{i}}{2R_{B}}$.

The contribution $U_{stretch}$ governs the elastic energy of the configuration III as follows:

\begin{equation}
\label{e.8}
U_{III} =  U_{0}+U_{fold}(h_{i},e_{c})+U_{inversion}(h_{i},e_{i})+U_{stretch}(h_{i},e_{i})
\end{equation}

Here, $U_{fold}(h_{i},e_{c})$ only varies by the thickness variation because $\ell$ is constant.

The energy $(U_{III}-U_{0})/(Eh^2\delta)$ is plotted as a function of the relative volume variation $\mid \frac{\Delta V}{V_{0}} \mid$ in figure \ref{fig:energy}: for small deformations, $U_{II} < U_{III}$ and configuration II is energetically more favourable than configuration III.
Note that this new configuration costs rapidly a high energy.

\begin{figure}
\centering
\resizebox{.8\columnwidth}{!}{
  \includegraphics{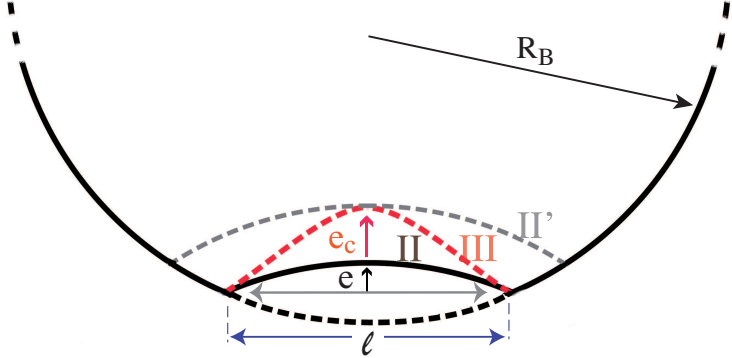}
}
\caption{Ideal scheme of the transition between different configurations of the elastic shell. Configuration II corresponds to the inversion of a part of the elastic shell characterized by an indentation length $e$ and a lateral size $\ell$. Configuration II can be followed by two possible configurations: (i) configuration II' corresponds to the increase of both the lateral size and the indentation length; (ii) configuration III where the lateral size is the same than in configuration II and the indentation length is the same than in configuration II'; consequently a stretching process of the inverted part is needed.}
\label{fig:sketch}       
\end{figure}

Let us now predict the critical quantity $e=e_{c}$, and consequently $\ell=\ell_{c}$, at which the transition between configuration II and III arises.
Starting with a deformed shell characterized by an indentation depth $e$ (configuration II in the sketch figure \ref{fig:sketch}), an increase of the indentation length can result in two possible configurations:

\begin{itemize}
\item both the indentation depth $e$ and the lateral size $\ell$ increase in accordance with configuration II' in figure \ref{fig:sketch});
\item only the indentation length increases at fixed lateral size $\ell=\ell_{c}$ in accordance with configuration III in \ref{fig:sketch}: in this case a length change of the inverted region is needed.
\end{itemize}

At the threshold, the energy of configuration II' has to equal the energy of configuration III.
Assuming that the elastic energy due to the inversion, $U_{inversion}$, is dominated both by the energy due to the folds and the one due to the stretching process, we have:
$U_{fold}(h,e+e_{c}) = U_{fold}(h,e)+U_{stretch}(h,e+e_{c})$ using the notations in figure \ref{fig:sketch}.
Starting from configuration II with $e<<e_{c}$, it comes: $U_{fold}(h,e_{c}) \sim U_{stretch}(h,e_{c})$ (at the threshold).
Using relations \ref{e.5} and \ref{e.7}, it comes directly that $e_{c}$ is proportional to the thickness at the onset of buckling, $h_{B}$, with a prefactor only depending on the Poisson ratio: $e_{c} = 8\frac{c_{0}}{c_{2}}h_{B}$.
Taking into account numerical values of the parameters $c_{0}$ and $c_{2}$ with a Poisson ratio equal $0.3$ we obtain\cite{Audoly10}: $e_{c} \sim 4h_{B}$.
In addition, since the transition to lower-energy results in a constant lateral size $\ell_{c}$, this length can be expressed as a function of the characteristics of the shell at the buckling onset as: $\ell_{c} \sim 2 (2R_Be_{c})^{1/2} = 4\sqrt{2}(R_B h_{B})^{1/2}$.

As a consequence the thickness of the shell at the onset of buckling $h_{B}$, can be precisely deduced from the measurement of the distorted drop at a macroscopic scale since:

\begin{equation}
\label{e.9}
h_{B} \sim \frac{l_{c}^2}{32R_{B}}
\end{equation}

\begin{figure}
\centering
\resizebox{0.55\columnwidth}{!}{
  \includegraphics{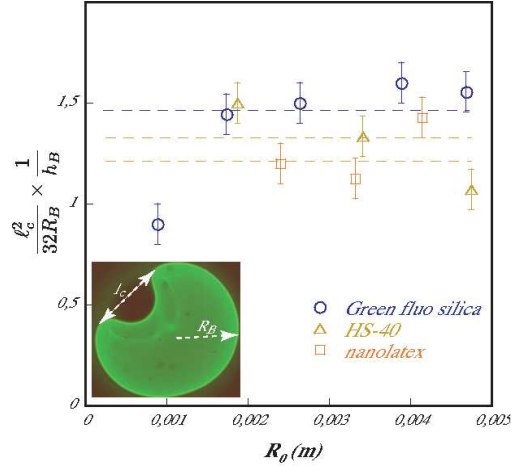}
  }
\caption{Measurements of the quantity $\frac{\ell_{c}^2}{32R_{B}} \times \frac{1}{h_{B}}$ for a range of initial drop radii $R_{0}$, and for different systems.}
\label{fig:lc}       
\end{figure}

Figure \ref{fig:lc} shows the dimensionless quantity $\frac{l_{c}^2}{32R_{B}}\times \frac{1}{h_{B}}$ for a range of initial drop radii.
Evaluation of $h_{B}$ using the relation \ref{e.9} is related to the measurements of macroscopic quantities ($l_{c}$, $R_{B}$) and is consequently more accurate than the direct measurement of $h_{B}$.
Typically, the error by direct measurement of $h_{B}$ is $20\%$ while it is equal to $5\%$ in the measurement of $\frac{l_{c}^2}{32R_{B}}$.
Note that the discrepancy between theory and experiment could be due to the systematic discrepancy between measurements of the enveloppe thickness and prediction as shown in figure \ref{fig:shell}d.

\begin{figure}
\centering
\resizebox{0.55\columnwidth}{!}{
  \includegraphics{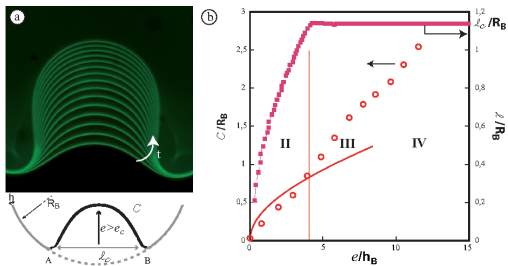}
}
\caption{The indentation depth $e$ at fixed lateral size $\ell=\ell_{c}$ turns into an invagination tip (superposition of profiles in (a)). 
(b) In configuration II, the length of the inverted cap, $C$, increases with the indentation depth $e$; the black line is the theoretical arc length of inverted cap of radius $R_{B}$ indented by $e$ (arc length $=R_{B}cos^{-1}(1-4e/R_{B})$).
In addition the lateral size $\ell$ increases with the indentation depth $e$.
The vertical line delimits configuration II and configuration III.
In configuration III, the lateral size keeps a constant value, $\ell=\ell_{c}$, while the length of the inverted part, $C$, still increases. When the length change of the inverted part reaches a critical value, possible decohesion in the inverted part occurs leading to a transition to configuration IV. This last configuration is characterized by the progression of an invagination tube that deepens in the droplet.}
\label{fig:invagination}       
\end{figure}

The continuation of the deformation is shown in figure \ref{fig:invagination}a: the depression deepens and forms an invaginating tube (configuration IV in figure \ref{fig:invagination}b).
As shown in figure \ref{fig:energy}, the stretching process of the shell is the main contribution to the elastic energy of configuration III.
However the energy required to proceed with this deformation is rapidly enormous.
Therefore a weakening or a breakage of the stretched part of the shell is suspected.

Indeed, a relative length change larger than $20\%$ as observed experimentally is often not feasible in such colloidal gels (figure \ref{fig:invagination}).
Let us compare the tensile stress in the stretched part of the shell with the cohesive stress in the shell.
The critical stress $\sigma_{c}$, needed to separate two particles of radius $a$ can be estimated from the van der Waals interaction in terms of equilibrium separation distance $Z_{0}$, and Hamaker constant $A$, whose value depends on the surface chemistry of the particles\cite{Israelachvili11}:
$\sigma_{c} \sim \frac{A a}{12 Z_{0}^2} \times a^{-2}$.

This critical stress is easily balanced by the tensile stress in the inverted region of the shell during the relative length change $u$: $E u \sim \sigma_{c}$.
Taking orders of magnitude for the Young modulus, $E \sim 10^7$Pa, obtained by indentation testing, and $A=0.83\times10^{-20}$J,  $Z_{0} \sim 0.2$nm for Ludox HS-40, the relative length change needed to separate particles is found to be: $u=0.18$.
This value corresponds to a relative volume variation of $\mid \Delta V/V_{0} \mid \sim 0.52$ (note that the same order of magnitude is obtained for the different colloidal systems studied).
Therefore the strong stretching of the shell possibly causes rapidly decohesion of the gel as the indentation depth grows.
Then, breakage or weakening of the inverted region is possibly followed by a healing process of the shell due to particles accumulation during drying.

\subsection{Case of spheroidal shells}
\begin{figure}
\centering
\resizebox{0.55\columnwidth}{!}{
  \includegraphics{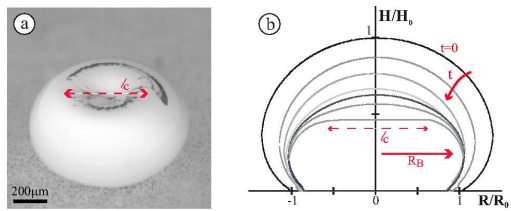}
}
\caption{Droplet of a Nanolatex dispersion deposited on a super-hydrophobic surface. (a) Distorted droplet: digitized image showing a depression of lateral size $l_{c}$ at the top of the droplet. (b) Superposition of dimensionless profiles (apex height $H$, vs. radius $R$) measured at different times by lateral imaging: the duration between two consecutive profiles is $300$ s.}
\label{fig:3D}      
\end{figure}

The case of three-dimensional droplets and large P\'eclet numbers corresponds to common industrial processes.
As in the confined geometry (2D case), drying a droplet of a colloidal dispersion leads to strong distortions (figure \ref{fig:3D}a).
However in 3D case, measurement of the shell thickness is not possible by direct observation.
The previous study can easily be applied to three-dimensional droplets to obtain shell thickness at the buckling instability.

The experiments were carried out with droplets of Nanolatex dispersion.
The nearly spheroidal geometry is obtained by depositing droplets on a hydrophobic substrate; the surface is sprayed with Lycopodium spores which form a rough texture in which air remains trapped when a drop is deposited.
In the first stage the droplet shrinks as described in the superposition of profiles in figure \ref{fig:3D}b.
Then an inverted region grows at the apex of the droplet (figure \ref{fig:3D}a,b)\cite{pauchard04}. 
As in the 2D case, the depression is continued by an invagination process.
Similar arguments can be applied to the case of spheroidal drops as explained in references \cite{pauchard97,pauchard98}. 
In particular the energy due to the fold and the one due to the stretching to the shell express now as:

\begin{equation}
\label{e.10}
U_{fold}^{3D}(h_{i},e{i}) = \frac{c'_{0}}{4} E \frac{h_{i}^{5/2}}{R_{B}} e_{i}^{3/2}
\end{equation}

\begin{equation}
\label{e.11}
U_{stretch}^{3D}(h_{i},e_{i}) =  \frac{\pi c'_{2}}{16} E \frac{h_{i}}{R_{B}} e_{i}^3
\end{equation}

$c'_{0}$ and $c'_{2}$ only depend on the Poisson ratio.
From \ref{e.10} and \ref{e.11} it comes that $e_{c}$ is proportional to the thickness at the onset of buckling $h_{B}$, such as:
$e_{c} \sim \big(\frac{4c'_{0}}{c'_{2}}\big)^{2/3}h_{B}$.

Since the geometric relation $\ell_{c} \sim 2\sqrt{2R_{B}e_{c}}$ still holds, it comes:

\begin{equation}
\label{e.12}
h_{B} \sim \frac{1}{8 \times 4^{2/3}} \frac{c'_{2}}{c'_{0}}^{2/3} \frac{l_{c}^2}{R_{B}} \sim \frac{1}{12}\frac{l_{c}^2}{R_{B}}
\end{equation}

Consequently, using relation \ref{e.12}, measurements of the macroscopic quantities $l_{c}$ and $R_{B}$ allow us to evaluate with a good accuracy the critical thickness for buckling of spheroidal shells.
In the case of a droplet of nanolatex dispersion (solid volume fraction $0.25$, particle diameter $25$ nm, glass transition temperature $=100^\circ$C) dried at $RH=50\%$ and $T=20^{\circ}C$, we find $h_{B}=25\pm2\mu$m. 

\section{Conclusion}
Desiccation of drops of complex fluids can display large shape distortions related to the development of an elastic skin at the drop surface.
We studied such mechanical instabilities in a confined geometry where direct measurements of the skin thickness are possible.
The drop deformation appears to be generic and does not depend on the initial drop size: a depression due to an inversion of curvature in the elastic shell is formed and deepens in the drop.
This deformation can be related to a mismatch between bending and stretching processes.
From the comparison between the corresponding elastic energies, we obtain a relation between the critical thickness for buckling and the characteristic lengths of the final drop shape that are easily measurable.
The validation of this method in the confined geometry (2D geometry) is then applied to spheroidal drops exhibiting identical distortions.

\section{Acknowledgment}
We would like to thank E. Sultan for useful discussions and A. Aubertin, L. Auffray, C. Borget, R. Pidoux for technical support.

\bibliography{biblio} 

\end{document}